\documentclass[aps,amsmath,amssymb,twocolumn,showpacs,prl]{revtex4-1}
\usepackage{graphicx}
\usepackage{bm}

\allowdisplaybreaks

\begin{document}

\title{Quantum-mechanical measurement apparatus as a black box}

\author{A.~V.~Nenashev}
\email {nenashev@isp.nsc.ru}
\affiliation{Rzhanov Institute of Semiconductor Physics, 630090 Novosibirsk, Russia}
\affiliation{Novosibirsk State University, 630090 Novosibirsk, Russia}

\date{\today}

\begin{abstract}
It is commonly believed that the most general type of a quantum-mechanical measurement is 
one described by a positive-operator valued measure (POVM). In the present paper, 
this statement is proven for any measurements on quantum systems with a finite-dimensional state space. 
The proof of POVM nature of an arbitrary measurement is carried out using a purely operational approach, 
which is fully ignorant about what is inside a measurement 
apparatus. 
The suggested approach gives also an opportunity to derive the Born rule.
\end{abstract}


\maketitle

In the early years of quantum theory, only measurements of a special kind were considered---namely, ones 
connected with \emph{observables}, which are Hermitian operators \cite{Dirac}. 
In the simplest case of 
a non-degenerate observable $\hat O$ and a pure state $|\psi\rangle$ of a measured system, 
the probability $P_O^{(k)}$ of getting the measurement result $O^{(k)}$ 
(one of eigenvalues of $\hat O$) 
is given by the Born rule:
\begin{equation} \label{eq:born-rule}
  P_O^{(k)} = \bigl| \langle\varphi_k|\psi\rangle \bigr|^2 ,
\end{equation}
where $|\varphi_k\rangle$ is the eigenvector corresponding to the eigenvalue $O^{(k)}$. More generally, 
$P_O^{(k)} = \mathrm{Tr} (\mathcal{\hat P}_O^{(k)} \hat\rho)$,
where $\hat\rho$ is the partial density matrix of the system under measurement, and 
$\mathcal{\hat P}_O^{(k)}$ is the projector onto the eigenspace of $\hat O$ with eigenvalue $O^{(k)}$. 
For this reason, measurements 
related to observables are sometimes called \emph{projective measurements} \cite{Nielsen}.

Later, it was recognized \cite{Davies1970} that there is a broader class of measurements, called 
\emph{general measurements} \cite{Nielsen}. A general measurement $M$ is characterized by a set 
$\{ \hat A_M^{(k)} \}$ 
of Hermitian operators, each operator corresponds to some ($k$th) outcome. The probability $P_M^{(k)}$ 
of getting the $k$th outcome is defined as 
\begin{equation} \label{eq:povm}
  P_M^{(k)} = \mathrm{Tr} (\hat A_M^{(k)} \hat\rho) .
\end{equation}
There are two requirements for the operators $\hat A_M^{(k)}$, following from Eq.~(\ref{eq:povm}) and 
properties of probability. The first one is non-negativity of their eigenvalues. 
The second one states than the sum $\sum_k \hat A_M^{(k)}$ is equal to the identity operator. 
A set of Hermitian operators obeying both requirements is usually called 
\emph{positive-operator valued measure} (POVM).

The question addressed in this paper is: are ``general measurements'' indeed \emph{general}? 
In other words: is it possible, for any given measurement apparatus $M$, to find such a POVM 
$\{ \hat A_M^{(k)} \}$ that probabilities $P_M^{(k)}$ of its outcomes will obey Eq.~(\ref{eq:povm}) 
for any state of a measured system?

There are several ways of introducing POVMs in quantum theory. POVMs occur in the case of 
\emph{indirect measurements}, when a system $A$ (to be measured) first interacts with another quantum 
system $B$, and actual (projective) measurement is then performed on the system $B$ 
\cite{Nielsen,Peres,Braginsky}. In this case, Eq.~(\ref{eq:povm}) follows from the Born rule~(\ref{eq:born-rule}). 
Also \emph{imperfect measurements}, where a result of a projective measurement is known to an observer up to 
some random error, can be naturally described in terms of POVMs \cite{Braginsky,Gardiner}. 
\emph{Continuous} and \emph{weak measurements} also lead to POVMs \cite{Jacobs2006}. 

These considerations, however, deal with particular cases of measurements, and therefore cannot provide 
an answer to the question on how general is the description of measurements by POVMs.
To get the answer, more suitable is an operational approach, 
where no assumptions are made about construction of a measurement apparatus, principle of its action, etc. 
Indeed, it has been shown \cite{Kraus,Holevo} 
that probabilities of outcomes $P_M^{(k)}$ of an arbitrary measurement $M$ obey 
Eq.~(\ref{eq:povm}) with an appropriately chosen POVM $\{ \hat A_M^{(k)} \}$, 
provided that \emph{these probabilities depend on 
the state of the measured system only through its density matrix} $\hat\rho$. 
(The statement given in italic will be referred to as ``assumption $\rho$'' below.)
Though assumption $\rho$ is usually accepted by default, its role should not be underestimated, 
because it contains some hidden statements about probabilities (see Ref. \onlinecite{Zurek2005}). 
For example, consider a measurement on an electron spin. Let $P_\uparrow$ and $P_\downarrow$ be probablilties 
of getting some outcome when the spin is up and down, respectively. Then, assumption $\rho$ implies that 
probability of this outcome must be equal to $(P_\uparrow+P_\downarrow)/2$ when the measured electron spin 
forms the singlet state $(|\!\!\uparrow\downarrow\rangle-|\!\!\downarrow\uparrow\rangle)/\sqrt2$ together with 
some other spin-1/2 particle. Such a strong restriction on values of probabilities needs justification. 
For this reason, in the present study we shall not require the measurement to satisfy assumption $\rho$.

The aim of this paper is to provide some thought experiments that prove Eq.~(\ref{eq:povm}) 
for arbitrary process of measurement. Our approach is fully operational. 
For illustrative purposes, we consider a measurement apparatus $M$ as being put in a black box (Fig.~1) 
that can accept some sort of particles 
(representing the measured quantum system) in its input. The only output of the black box is the lamp on it, 
which flashes for a moment each time when the measurement outcome is equal to some fixed value $k$. 

It is enough to prove Eq.~(\ref{eq:povm}) only for \emph{pure} 
states, as of the system under measurement, as of larger systems including some 
\emph{environment}. Once Eq.~(\ref{eq:povm}) is established for pure states, it can be easily 
generalized to the case of probabilistic mixtures, see Appendix A.

The proof of Eq.~(\ref{eq:povm}) for pure states will be given in two stages. At the first stage, we will ensure, 
by considering the experiments shown in Fig.~1, that for any two pure states $S$ and $S'$ having 
the same partial density matrix of the measured system, the probabilities of measurement outcomes 
are the same:
\begin{equation} \label{eq:same-rho-same-p}
  \hat \rho_S = \hat \rho_{S'}  \quad\Rightarrow\quad
  P_M^{(k)} (S) = P_M^{(k)} (S') .
\end{equation}
In other words, the probability $P_M^{(k)}$ is only a function of the partial density matrix $\hat\rho_S$ 
(for given measuring device $M$ and outcome $k$). We will denote this function as $F_M^{(k)}$:
\begin{equation} \label{eq:f}
  P_M^{(k)} (S) = F_M^{(k)} (\hat \rho_S) .
\end{equation}

At the second stage, we will use the thought experiments shown in Fig.~2 to 
prove that the function $F_M^{(k)}$ is \emph{linear}. More precisely, we will show that for any two density 
matrices $\hat\rho_0$ and $\hat\rho_1$ and any real number $\lambda\in[0,1]$ 
\begin{equation} \label{eq:f-linearity}
  F_M^{(k)} \bigl( (1-\lambda)\hat\rho_0 + \lambda\hat\rho_1 \bigl) = 
  (1-\lambda) \, F_M^{(k)} (\hat\rho_0) + \lambda \, F_M^{(k)} (\hat\rho_1) .
\end{equation}
Such linearity gives the possibility to express the function $F_M^{(k)}$ in the following form:
\begin{equation} \label{eq:f-trace}
  F_M^{(k)} (\hat\rho) = \mathrm{Tr} (\hat A_M^{(k)} \hat\rho) ,
\end{equation}
where $\hat A_M^{(k)}$ is some non-negative Hermitian operator. 
Substitution of Eq.~(\ref{eq:f-trace}) into Eq.~(\ref{eq:f}) gives Eq.~(\ref{eq:povm}), that completes the proof 
of the POVM nature of an arbitrary quantum-mechanical measurement.

We restrict ourselves in this paper by consideration only measurements on \emph{finite-dimensional} 
systems.

For a better clarity, let us schematically picture a typical quantum-mechanical experiment as shown in Fig.~1a. 
At pressing the button, the source $S$ emits two particles: $A$ and $B$. The particle $A$ represents the 
system to be measured, and the particle $B$ plays the role of an environment to which the particle $A$ might be 
entangled. The source prepares the composite system of two particles in a pure state $|\Psi\rangle$, 
which stays unchanged until the particle $A$ reaches the measuring apparatus $M$.
This apparatus is equipped with a lamp that flashes when the measurement gives the result $k$. 
An observer is sitting near the measuring device $M$ and is counting the frequency of flashing the lamp. 
This frequency, being divided by the frequency of emitting the pairs of particles by the source $S$, gives the 
probability $P_M^{(k)} (S)$.
\begin{figure}
\includegraphics[width=\linewidth]{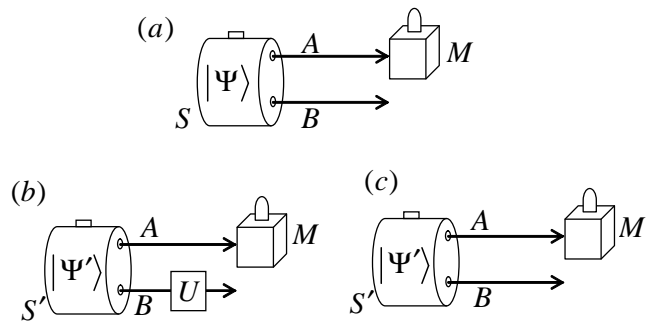}
\caption{Three thought experiments used in the proof of Eq.~(\ref{eq:same-rho-same-p}). 
The source $S$ ($S'$) emits particles $A$ and $B$ prepared in the joint state $\Psi$ ($\Psi'$). 
The particle $A$ then reaches the measuring apparatus $M$. The lamp on the apparatus $M$ flashes 
when the measurement gives the result $k$. 
In the experiment ({\it b}), the particle $B$ undergoes the transformation $\hat U$ defined by Eq.~(\ref{eq:u}).
}
\label{fig:1}
\end{figure}

Any pure state $|\Psi\rangle$ of a system of two particles ($A$ and $B$) can be represented in the form of 
the Schmidt decomposition:
\begin{equation} \label{eq:schmidt-psi}
  |\Psi\rangle = \sum_{n=1}^N c_n |\varphi_n\rangle |\chi_n\rangle ,
\end{equation}
where $N$ is the smallest of dimensionalities of the two particles' state spaces; 
$c_n$ are non-negative real numbers; $|\varphi_n\rangle$ are mutually orthogonal unit vectors 
in the state space of the particle $A$; and $|\chi_n\rangle$ are mutually orthogonal unit vectors 
in the state space of the particle $B$:
\[
  \langle\varphi_m|\varphi_n\rangle = \langle\chi_m|\chi_n\rangle = \delta_{mn} ,
\]
$\delta_{mn}$ being the Kroneker's delta.

The partial density matrix of the particle $A$ for the state $|\Psi\rangle$ is 
\begin{equation} \label{eq:rho-psi}
  \hat\rho = \sum_{n=1}^N c_n^2 |\varphi_n\rangle \langle\varphi_n| .
\end{equation}
It does not depend on the vectors $|\chi_n\rangle$. Consequently, any state $|\Psi'\rangle$ having the 
Schmidt decomposition 
\begin{equation} \label{eq:schmidt-psi2}
  |\Psi'\rangle = \sum_{n=1}^N c_n |\varphi_n\rangle |\chi'_n\rangle 
\end{equation}
with the same sets of numbers $c_n$ and vectors $|\varphi_n\rangle$ as in the 
decomposition~(\ref{eq:schmidt-psi}), 
but with a different set of mutually orthogonal unit vectors $|\chi'_n\rangle$, 
has the same partial density matrix of the particle $A$ as for the state $|\Psi\rangle$.

It is easy to show that the converse statement is also true (see Appendix B). 
Namely, if two different pure states $|\Psi\rangle$ and $|\Psi'\rangle$ of the bipartite system have the same 
partial density matrix of the particle $A$, then their Schmidt decompositions can be chosen in the 
forms~(\ref{eq:schmidt-psi}) and~(\ref{eq:schmidt-psi2}), with the same sets of $c_n$ and $|\varphi_n\rangle$. 
As both sets $|\chi_n\rangle$ and $|\chi'_n\rangle$ are orthonormal (by definition of the Schmidt decomposition), 
there is some unitary operator $\hat U$ in the state space of the particle $B$ that maps 
the set $|\chi'_n\rangle$ into the set $|\chi_n\rangle$:
\begin{equation} \label{eq:u}
  \forall\, n=1,\ldots,N  \qquad  \hat U |\chi'_n\rangle = |\chi_n\rangle .
\end{equation}
Such an unitary operator $\hat U$ can be implemented (at least in a thought experiment) as a physical device 
that performs the transformation $\hat U$ upon the particle $B$.

Now let us consider the experiment depicted in Fig.~1b. The source $S'$ prepares a pair of particles 
($A$ and $B$) in a pure state $|\Psi'\rangle$, for which the partial density matrix of the particle $A$ 
is the same as for the state $|\Psi\rangle$. Then the particle $B$ passes through a device that implements 
the operator $\hat U$ introduced in Eq.~(\ref{eq:u}), where vectors $|\chi_n\rangle$ and $|\chi'_n\rangle$ 
are defined by Eqs.~(\ref{eq:schmidt-psi}) and~(\ref{eq:schmidt-psi2}). After that, the particle $A$ reaches 
the measuring apparatus $M$. 
Just before the measurement, a joint state of the particles $A$ and $B$ is
\[
  \sum_{n=1}^N c_n |\varphi_n\rangle (\hat U |\chi'_n\rangle) = 
  \sum_{n=1}^N c_n |\varphi_n\rangle |\chi_n\rangle   \equiv   |\Psi\rangle ,
\]
i.~e. the same as in the experiment shown in Fig.~1a. Consequently, there is no difference between frequencies 
of lamp flashing in the two experiments shown in Figs.~1a and~1b. (We imply that there are no hidden variables, 
i.~e. the state vector fully determines all statistics.)

Also this frequency will not change if the device performing the operation $\hat U$ is removed (Fig.~1c). 
This is because there is no causal link (no interaction) between particles $A$ and $B$ after they left the 
source $S'$; as a consequence, no information about the fate of the particle $B$ is available during 
the measurement.

Thus, if the sources $S$ and $S'$ produce the same partial density matrix of the particle $A$ 
($\hat\rho_S = \hat\rho_{S'}$), then the probabilities of lamp flashing in the experiments of Fig.~1a and 
of Fig.~1c will be the same: $P_M^{(k)} (S) = P_M^{(k)} (S')$. This proves Eq.~(\ref{eq:same-rho-same-p}) and, 
consequently, Eq.~(\ref{eq:f}).

We presumed above that \emph{the same} particle $B$ plays the role of an environment 
in the states $S$ and $S'$. 
It is possible to show (see Appendix C) that 
Eq.~(\ref{eq:same-rho-same-p}) stays in force even in the case of different environments, 
which makes the function $F_M^{(k)}$ independent of the kind of environment.

The next step is to prove linearity of the function $F_M^{(k)}$, Eq.~(\ref{eq:f-linearity}). 
Let $\hat\rho_0$ and $\hat\rho_1$ be two arbitrarily chosen density matrices of some particle $A$. 
One can always choose such a particle $B$ and such two \emph{pure} states $|\Psi_0\rangle$ and $|\Psi_1\rangle$ 
of the sysyem of two particles $A$ and $B$, that the reduced density matrix of the particle $A$ 
is equal to $\hat\rho_0$ for the state $|\Psi_0\rangle$, and to $\hat\rho_1$ for the state $|\Psi_1\rangle$. 
Then, let us consider a thought experiment shown in Fig.~2a. The source $S_0$ emits a pair of 
particles $A$ and $B$ prepared in the state $|\Psi_0\rangle$. Simultaneously, another source $Q_\lambda$ 
emits a pair of entangled \emph{qubits} (e.~g. spin-1/2 particles) $\alpha$ and $\beta$ in the state
\begin{equation} \label{eq:phi-lambda}
  |\Phi_\lambda\rangle = \sqrt{1-\lambda}\, |0\rangle |0\rangle + \sqrt{\lambda}\, |1\rangle |1\rangle ,
\end{equation}
where $\lambda$ is an adjustable parameter, $0\leq\lambda\leq1$. The joint state of four particles 
$A,B,\alpha,\beta$ is therefore equal to
\[
  |\Psi_0\rangle |\Phi_\lambda\rangle  \equiv  
  \sqrt{1-\lambda}\, |\Psi_0\rangle |0\rangle |0\rangle + \sqrt{\lambda}\, |\Psi_0\rangle |1\rangle |1\rangle .
\]
Then three particles $A$, $B$ and $\alpha$ go through a ``quantum gate'' $G$ that performs the following 
``controlled transformation'':
\begin{align} 
  \label{eq:g0}
  |\Psi_0\rangle |0\rangle  &\stackrel{G}{\to}  |\Psi_0\rangle |0\rangle , \\
  \label{eq:g1}
  |\Psi_0\rangle |1\rangle  &\stackrel{G}{\to}  |\Psi_1\rangle |1\rangle ,
\end{align}
i.~e. if the qubit $\alpha$ is in the state $|0\rangle$, then nothing will be changed; if it is in the 
state $|1\rangle$, then the system $A+B$ will undergo an unitary transformation which maps the vector 
$|\Psi_0\rangle$ onto the vector $|\Psi_1\rangle$.
The state of the four particles after the gate $G$ is
\[
  \sqrt{1-\lambda}\, |\Psi_0\rangle |0\rangle |0\rangle + \sqrt{\lambda}\, |\Psi_1\rangle |1\rangle |1\rangle .
\]
For this state, the reduced density matrix $\hat\rho$ of the particle $A$ is
\begin{equation} \label{eq:rho-after-g}
  \hat\rho = (1-\lambda)\hat\rho_0 + \lambda\hat\rho_1 .
\end{equation}
Finally, the particle $A$ is measured by the same apparatus $M$ that was considered above. We are interested in 
the probability $p(M:\text{flash})$ that the lamp on the apparatus $M$ will flash. According to 
Eqs.~(\ref{eq:f}) and~(\ref{eq:rho-after-g}),
\begin{equation} \label{eq:p-m-flash-a}
  p(M:\text{flash}) = F_M^{(k)} \bigl( (1-\lambda)\hat\rho_0 + \lambda\hat\rho_1 \bigl) .
\end{equation}
\begin{figure}
\includegraphics[width=0.9\linewidth]{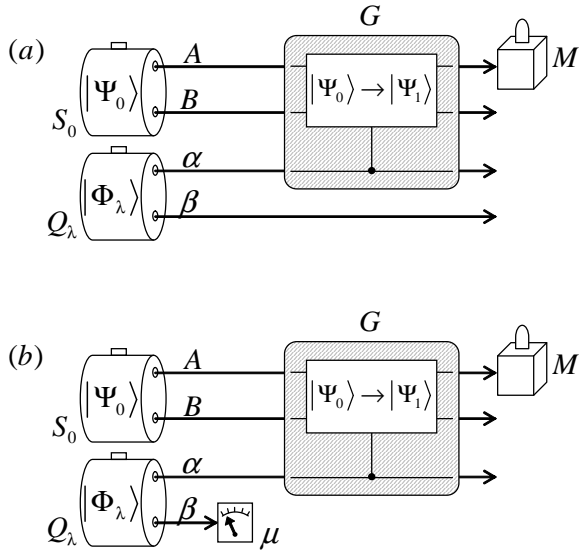}
\caption{Thought experiments used in the proof of Eq.~(\ref{eq:f-linearity}). 
The source $S_0$ emits particles $A$ and $B$ prepared in a joint pure state $|\Psi_0\rangle$. Simultaneously, 
another source $Q_\lambda$ emits a pair of entangled \emph{qubits} $\alpha$ and $\beta$ 
prepared in the state $|\Phi_\lambda\rangle$ defined by Eq.~(\ref{eq:phi-lambda}). 
Then, particles $A$, $B$ and $\alpha$ pass through a quantum gate $G$ that operates according to 
Eqs.~(\ref{eq:g0}) and~(\ref{eq:g1}). After passing through the gate $G$, the particle $A$ reaches the 
measurement apparatus $M$. The lamp on $M$ flashes when the measurement outcome is equal to $k$. 
In the experiment $b$, the qubit $\beta$ is measured in the basis 
$\{ |0\rangle,|1\rangle \}$ by the meter $\mu$ before the particles $A$, $B$, $\alpha$ reach the gate $G$.
}
\label{fig:2}
\end{figure}

Now we will consider a modification of this experiment shown in Fig.~2b. The only difference between 
Figs.~2a and~2b is that, in the latter experiment, the qubit $\beta$ is measured in the basis 
$\{ |0\rangle,|1\rangle \}$ (by the ``meter'' $\mu$) before the rest three particles reach the gate $G$. 

In both experiments, the trajectory of the particle $\beta$, together with the meter $\mu$, is 
spatially separated from (and is not interacting to) the rest of the setup, therefore no information 
about this particle can reach the measuring device $M$. Consequently the probability $p(M:\text{flash})$ 
is the same for both experiments. For the second experiment (Fig.~2b), one can apply the law of 
total probability to the quantity $p(M:\text{flash})$:
\begin{align} \label{eq:total-probability}
  p(M:\text{flash}) 
  &= p(\mu:0) \, p(M:\text{flash}\,|\,\mu:0) \nonumber \\
  &+ p(\mu:1) \, p(M:\text{flash}\,|\,\mu:1) ,
\end{align}
where $p(\mu:x)$ is the probability that the meter $\mu$ will give the result $x$ ($x$ is either 0 or 1); 
$p(M:\text{flash}\,|\,\mu:x)$ is the \emph{conditional} probability of lamp flashing on the device $M$ 
provided that the meter $\mu$ gives the result $x$. 
The value of $p(\mu:1)$ does not depend on the state $|\Psi_0\rangle$, but depends on choice of $\lambda$. 
Let us denote this quantity as $a_\lambda$:
\footnote{Applying the Born rule to the state vector $\Phi_\lambda$, one can immediately find that 
$a_\lambda = \lambda$. However, we will avoid using the Born rule in this paper, and will prove the 
equality $a_\lambda = \lambda$ in another way. It will give us a possibility to \emph{derive} the Born rule, 
see the discussion below.}
\begin{equation} \label{eq:a-lambda}
  p(\mu:1) = a_\lambda\, ,  \quad  p(\mu:0) = 1-a_\lambda\, .
\end{equation}
By definition of $a_\lambda$,
\begin{gather} 
  \label{eq:a-lambda-restriction}
  0 \leq a_\lambda \leq 1 , \\
  \label{eq:a-lambda-01}
  a_0=0 , \quad a_1=1 .
\end{gather}

If the meter $\mu$ gives the result 0, then the qubit 
$\alpha$ will appear in the state $|0\rangle$ after the measurement of the qubit $\beta$, 
due to the perfect correlation between the two entangled qubits in the state $|\Phi_\lambda\rangle$. 
According to Eq.~(\ref{eq:g0}), in this case the particles $A$ and $B$ will remain to be in the state 
$|\Psi_0\rangle$ after passing through the gate $G$. Thus, the partial density matrix of the particle $A$ 
before its measurement will be equal to $\hat\rho_0$, and
\begin{equation} \label{eq:p-m-flash-mu0}
  p(M:\text{flash}\,|\,\mu:0) = F_M^{(k)}(\hat\rho_0) .
\end{equation}
Similarly, if the result of measurement the qubit $\beta$ is 1, then the qubit $\alpha$ will be in the 
state $|1\rangle$ after this measurement. In this case, the gate $G$ will change the state of particles 
$A$ and $B$ from $|\Psi_0\rangle$ to $|\Psi_1\rangle$, according to Eq.~(\ref{eq:g1}), and 
the partial density matrix of the particle $A$ before its measurement will be equal to $\hat\rho_1$. Hence,
\begin{equation} \label{eq:p-m-flash-mu1}
  p(M:\text{flash}\,|\,\mu:1) = F_M^{(k)}(\hat\rho_1) .
\end{equation}
Substituting the results~(\ref{eq:a-lambda}), (\ref{eq:p-m-flash-mu0}), and~(\ref{eq:p-m-flash-mu1}) 
into Eq.~(\ref{eq:total-probability}), and comparing with Eq.~(\ref{eq:p-m-flash-a}), one can obtain 
the following relation between probabilities~$F_M^{(k)}$ for different density matrices:
\begin{equation} \label{eq:f-linearity-a-lambda}
  F_M^{(k)} \bigl( (1-\lambda)\hat\rho_0 + \lambda\hat\rho_1 \bigl) = 
  (1-a_\lambda) \, F_M^{(k)} (\hat\rho_0) + a_\lambda \, F_M^{(k)} (\hat\rho_1) .
\end{equation}
This equation is valid for any possible density matrices $\hat\rho_0$, $\hat\rho_1$ of the particle $A$, 
and for any values of $\lambda\in[0,1]$. 

Eq.~(\ref{eq:f-linearity-a-lambda}), together with conditions~(\ref{eq:a-lambda-restriction}) 
and~(\ref{eq:a-lambda-01}), provides enough background to prove that
\begin{equation} \label{eq:a-lambda-is-lambda}
  \forall \lambda\in[0,1]  \quad  a_\lambda = \lambda .
\end{equation}
For a proof of Eq.~(\ref{eq:a-lambda-is-lambda}), see Appendix D. 
Substitution $a_\lambda = \lambda$ in Eq.~(\ref{eq:f-linearity-a-lambda}) completes the proof of
Eq.~(\ref{eq:f-linearity}). 

Now we will show how Eq.~(\ref{eq:f-trace}) follows from Eq.~(\ref{eq:f-linearity}). 
Let us consider a density matrix $\hat\rho$ as a point in the real space, whose coordinates are 
real and imaginary parts of the matrix elements $\rho_{mn}$. The set $\Omega$ of all density matrices 
is a convex subset of this real space. According to 
Eq.~(\ref{eq:f-linearity}), the function $F_M^{(k)}(\hat\rho)$ is linear on any line segment inside $\Omega$. 
Hence, this function is linear over the whole set $\Omega$. As shown in Ref.~\onlinecite{Holevo},  
Lemma 1.6.2 (see also Appendix E), 
any such a linear function has the form $\mathrm{Tr} (\hat A \hat\rho)$ with an appropriate 
Hamiltonian operator $\hat A$. This justifies Eq.~(\ref{eq:f-trace}). 

Finally, Eq.~(\ref{eq:f-trace}) together with Eq.~(\ref{eq:f}) gives 
Eq.~(\ref{eq:povm}), proving thereby the statement that any measurement in quantum mechanics can be 
described by POVM.

It should be noted that the presented derivation of Eq.~(\ref{eq:povm}) uses neither 
the Born rule~(\ref{eq:born-rule}), nor any other form of quantum-mechanical probabilistic postulate. 
This opens the possibility to \emph{derive} the Born rule from Eq.~(\ref{eq:povm}).  
Such a possibility is demonstrated in Appendix F for 
the case of \emph{maximal} measurement, i.~e. when the number of possible outcomes is equal to the 
dimensionality $N$ of the state space of the measured system. 
If there are such $N$ states $S_k$, that for each of them the measuring apparatus $M$ 
gives the corresponding ($k$th) outcome \emph{with certainty}, then:
\newline
(i) each state $S_k$ is a \emph{pure} state of the measured system;
\newline
(ii) state vectors $|\varphi_k\rangle$ corresponding to the states $S_k$ are mutually orthogonal: 
$\langle\varphi_k| \varphi_l\rangle = \delta_{kl}$;
\newline
(iii) the probability $P_M^{(k)} (S)$ of $k$th outcome for an arbitrary state $S$ is equal to 
$\langle\varphi_k| \hat\rho_S |\varphi_k\rangle$, which gives the Born rule~(\ref{eq:born-rule}) in the case 
of measurement of pure states.

There are many other ways of deriving the Born rule 
\cite{Gleason1957,Everett1957,Hartle1968,Valentini1991a,Deutsch1999,Hardy2001,Zurek2003,Rubin2003,
Busch2003,Caves2004,Saunders2004,Zurek2005,Wallace2009} 
(for review, see Refs.~\onlinecite{Schlosshauer2005,Dickson2011}); each of them starts 
from its own set of axioms. 
The starting point of our approach is roughly similar to that of Zurek's ``envariance'' 
(i.~e. environment-induced envariance) method~\cite{Zurek2003,Zurek2005}, 
and of Saunders' operational method~\cite{Saunders2004}. 
The advantage of our approach is its simplicity 
(all its essence is pictured in Figs.~1,2) and its broader scope (applicability to both projective and 
non-projective measurements).

We emphasize that \emph{orthogonality} of state vectors corresponding to different outcomes of a projective 
measurement \emph{can be derived} by our method, rather than postulated. An ultimate reason for this 
orthogonality is the unitary (norm-conserving) dynamics of quantum-mechanical systems between their 
preparation and measurement.

Entanglement plays a key role in our approach. 
Importance of entanglement for justification of the probability rule 
has been emphasized by Zurek, who obtained the Born rule considering the symmetries of maximally 
entangled states \cite{Zurek2003,Zurek2005}. 
Our method can be viewed as a generalization of the Zurek's method of ``envariance'' 
to the case of general measurements.

In conclusion, we have answered 
(by means of thought experiments shown in Figs.~1,2)
to the following question: what is the most general type of probability 
rule in quantum-mechanical measurements, irrespective to internal structure and operation principle of 
a measurement device? We have shown that, under reasonable assumptions,  
any possible measurement is described by a POVM, i.~e. probabilities of its outcomes obey Eq.~(\ref{eq:povm}). 
These assumptions are: 
\begin{itemize}
\item the Hilbert space formalism for state vectors; 
\item the possibility of preparing any pure state and of performing any unitary transformation; 
\item no hidden variables; 
\item the law of total probability for macroscopic events (e.~g. measurement outcomes); 
\item the perfect correlation between two entangled qubits prepared in the state 
$\sqrt{1-\lambda}\, |0\rangle |0\rangle + \sqrt{\lambda}\, |1\rangle |1\rangle$; 
\item and impossibility of information transfer without interaction. 
\end{itemize}

% ---------------------------------------------

\newpage
\subsection{Appendix A. Generalization of the probability rule to the case of mixed states}

Let us consider a composite system $A+B$, a part $A$ of which is to be measured, and another part $B$ 
plays the role of an environment. Let $P(S)$ be the probability that, 
for the state $S$ of the system $A+B$, measurement on the part $A$ by some apparatus $M$ will give 
the $k$-th outcome. Suppose that the dependence of $P(S)$ has the form
\begin{equation} \label{a:povm}
  P(S) = \mathrm{Tr} (\hat A \hat\rho_S) 
\end{equation}
for any \emph{pure} state $S$, where $\hat\rho_S$ is the partial density matrix of the system $A$ for the 
state $S$, and $\hat A$ is some Hermitian matrix.

In this Section, we will show that Eq.~(\ref{a:povm}) can be generalized to the case of \emph{mixed} states.

A mixed state $\mathfrak{M}$ of the system $A+B$ can be considered as a 
collection $S_1,S_2,\ldots,S_L$ of pure states of this system; each pure state $S_l$ appears with its 
corresponding probability $p_l$. Hence, one can apply the law of total probability:
\[
  P(\mathfrak{M}) = \sum_{l=1}^L p_l \, P(S_l) .
\]
Then, let us use Eq.~(\ref{a:povm}) for evaluating the probabilities $P(S_l)$:
\begin{equation} \label{a:p-mixed}
  P(\mathfrak{M}) = \sum_{l=1}^L p_l \, \mathrm{Tr} ( \hat A \hat\rho_l ) = 
  \mathrm{Tr} \left( \hat A \sum_{l=1}^L p_l \, \hat\rho_l \right) ,
\end{equation}
where $\hat\rho_l$ is the density matrix for the state $S_l$. 
Introducing the partial density matrix $\hat\rho_{\mathfrak{M}}$ of the system $A$ for the mixed state 
$\mathfrak{M}$,
\[
  \hat\rho_{\mathfrak{M}} = \sum_{l=1}^L p_l \, \hat\rho_l ,
\]
one can rewrite Eq.~(\ref{a:p-mixed}) as 
\begin{equation} \label{a:povm-mixed}
  P(\mathfrak{M}) = \mathrm{Tr} (\hat A \hat\rho_{\mathfrak{M}}) .
\end{equation}
The latter equation generalizes Eq.~(\ref{a:povm}) to the case of mixed states.

\subsection{Appendix B. Similarity of Schmidt decompositions of two states having the same partial density matrix}

Any pure state $|\Psi\rangle$ of a system of two parts $A$ and $B$ can be represented in the form of 
the Schmidt decomposition:
\begin{equation} \label{b:schmidt-psi}
  |\Psi\rangle = \sum_{n=1}^N c_n |\varphi_n\rangle |\chi_n\rangle ,
\end{equation}
where $N$ is the smallest of dimensionalities of the two parts' state spaces; 
$c_n$ are non-negative real numbers; $|\varphi_n\rangle$ are mutually orthogonal unit vectors 
in the state space of the part $A$; and $|\chi_n\rangle$ are mutually orthogonal unit vectors 
in the state space of the part $B$:
\[
  \langle\varphi_m|\varphi_n\rangle = \langle\chi_m|\chi_n\rangle = \delta_{mn} ,
\]
$\delta_{mn}$ being the Kroneker's delta.

In this Section, we will show that if another pure state $|\Psi'\rangle$ of the same system $A+B$ 
has the same partial density matrix of the part $A$ as the state $|\Psi\rangle$, then the Schmidt 
decomposition of the vector $|\Psi'\rangle$ can be chosen as
\begin{equation} \label{b:schmidt-psi2}
  |\Psi'\rangle = \sum_{n=1}^N c_n |\varphi_n\rangle |\chi'_n\rangle ,
\end{equation}
i.~e. with the same sets of coefficients $c_n$ and of part $A$'s vectors $|\varphi_n\rangle$, and 
with some orthonormal set of part $B$'s vectors $|\chi'_n\rangle$:
\begin{equation} \label{b:chi2-orthonormal}
  \langle\chi'_m|\chi'_n\rangle = \delta_{mn} .
\end{equation}
The vectors $|\Psi\rangle$ and $|\Psi'\rangle$ are supposed to be normalized.

For simplicity, we will consider the case when both parts ($A$ and $B$) have the same dimensionality $N$ 
of their state spaces. Generalization to the case of different dimensionalities is straightforward.

To prove possibility of the Schmidt decomposition~(\ref{b:schmidt-psi2}), we will start from an arbitrary 
Schmidt decomposition of $|\Psi'\rangle$, 
\begin{gather} \label{b:schmidt-psi2-arbitrary}
  |\Psi'\rangle = \sum_{n=1}^N \tilde c_n |\tilde\varphi_n\rangle |\tilde\chi_n\rangle , \\
  \text{where} \quad \tilde c_n \geq 0, \quad
  \langle\tilde\varphi_m|\tilde\varphi_n\rangle = \langle\tilde\chi_m|\tilde\chi_n\rangle = \delta_{mn} ,
\end{gather}
and will construct the set of vectors $|\chi'_n\rangle$ that satisfy 
Eqs.~(\ref{b:schmidt-psi2}) and~(\ref{b:chi2-orthonormal}).

The partial density function of the part $A$ for the state $|\Psi\rangle$ is 
\begin{equation} \label{b:rho-psi}
  \hat\rho = \sum_{n=1}^N c_n^2 |\varphi_n\rangle \langle\varphi_n| .
\end{equation}
One can see from this equation that the coefficients $c_n$ are square roots of eigenvalues of the 
density matrix $\hat\rho$. Since the density matrix is the same for vectors $|\Psi\rangle$ and $|\Psi'\rangle$, 
the set of coefficients $c_n$ is the same as the set of $\tilde c_n$. One can therefore assume, without any 
loss of generality, that
\begin{equation} \label{b:c-n-the-same}
  \tilde c_n = c_n .
\end{equation}
Also it can be seen form Eq.~(\ref{b:rho-psi}) that each vector $|\varphi_n\rangle$ is an eigenvector of the 
matrix $\hat\rho$ with the corresponding eigenvalue $c_n^2$. The same is true for vectors 
$|\tilde\varphi_n\rangle$. Eigenvectors corresponding to non-equal eigenvalues are mutually orthogonal; 
consequently, if $c_n \neq c_m$ then $\langle\varphi_m|\tilde\varphi_n\rangle = 0$. This statement can be 
expressed as follows:
\begin{equation} \label{b:c-n-to-c-m}
  c_n \langle\varphi_m|\tilde\varphi_n\rangle =  
  c_m \langle\varphi_m|\tilde\varphi_n\rangle ,
\end{equation}

Now let us write down the expansion of vectors $|\tilde\varphi_n\rangle$ in the basis of vectors 
$|\varphi_m\rangle$,
\[
  |\tilde\varphi_n\rangle = \sum_{m=1}^N  |\varphi_m\rangle  \langle\varphi_m|\tilde\varphi_n\rangle ,
\]
and substitute this expansion into Eq.~(\ref{b:schmidt-psi2-arbitrary}), taking also into account that 
$\tilde c_n = c_n$:
\begin{equation} \label{b:psi2-expansion}
  |\Psi'\rangle = 
  \sum_{m=1}^N \sum_{n=1}^N c_n |\varphi_m\rangle \langle\varphi_m|\tilde\varphi_n\rangle |\tilde\chi_n\rangle .
\end{equation}
Due to Eq.~(\ref{b:c-n-to-c-m}), one can change factors $c_n$ by $c_m$ in Eq.~(\ref{b:psi2-expansion}), 
yielding
\begin{multline}  \label{b:psi2-expansion2}
  |\Psi'\rangle = 
  \sum_{m=1}^N \sum_{n=1}^N c_m |\varphi_m\rangle \langle\varphi_m|\tilde\varphi_n\rangle |\tilde\chi_n\rangle 
  \\ =
  \sum_{m=1}^N c_m |\varphi_m\rangle 
  \left( \sum_{n=1}^N \langle\varphi_m|\tilde\varphi_n\rangle |\tilde\chi_n\rangle \right) .
\end{multline}
Finally, considering the expressions in brackets in Eq.~(\ref{b:psi2-expansion2}) 
as the sought-for vectors $|\chi'_m\rangle$,
\[
  |\chi'_m\rangle  =  \sum_{n=1}^N \langle\varphi_m|\tilde\varphi_n\rangle |\tilde\chi_n\rangle ,
\]
we arrive to the equality
\[
  |\Psi'\rangle = \sum_{m=1}^N c_m |\varphi_m\rangle |\chi'_m\rangle ,
\]
which is equivalent to Eq.~(\ref{b:schmidt-psi2}). Thus, Eq.~(\ref{b:schmidt-psi2}) is justified.

The last thing to do is checking Eq.~(\ref{b:chi2-orthonormal}), which is straightforward:
\begin{multline*}
  \langle\chi'_n|\chi'_m\rangle = 
  \left( \sum_{a=1}^N \langle\tilde\varphi_a|\varphi_n\rangle \langle\tilde\chi_a| \right) 
  \left( \sum_{b=1}^N \langle\varphi_m|\tilde\varphi_b\rangle |\tilde\chi_b\rangle \right) 
  \\ 
  = \sum_{a=1}^N \sum_{b=1}^N \langle\tilde\chi_a|\tilde\chi_b\rangle 
  \langle\varphi_m|\tilde\varphi_b\rangle \langle\tilde\varphi_a|\varphi_n\rangle 
  \\ 
  = \sum_{a=1}^N \langle\varphi_m|\tilde\varphi_a\rangle \langle\tilde\varphi_a|\varphi_n\rangle 
  = \langle\varphi_m|\varphi_n\rangle = \delta_{mn} .
\end{multline*}

\subsection{Appendix C. The case of different environments}

In the discussion of Fig.~1 (see the main article), we considered such two pure states 
$|\Psi\rangle$ and $|\Psi'\rangle$ of some composite system $A+B$, that the partial density matrix 
of the subsystem $A$ is the same for $|\Psi\rangle$ and for $|\Psi'\rangle$. We had shown that 
probability of any outcome of any measurement on $A$ has the same value for the 
system $A+B$ prepared in the state $|\Psi\rangle$ and in the state $|\Psi'\rangle$.

Now we will generalize this statement to the case when $|\Psi\rangle$ and $|\Psi'\rangle$ are 
states of \emph{different} composite systems. Let us denote these systems as $\mathbb{S}$ and $\mathbb{S}'$. 
Both $\mathbb{S}$ and $\mathbb{S}'$ include $A$ as a subsystem. Besides $A$, the systems 
$\mathbb{S}$ and $\mathbb{S}'$ can share some other common part; let us denote it as $B$. 
In a general case, one can therefore represent the system $\mathbb{S}$ as a combination $A+B+C$, 
and the system $\mathbb{S}'$ as $A+B+C'$, where subsystems $C$ and $C'$ have no intersections.

Let us consider four experiments shown in Fig.~3: 
\newline
(a) preparation of the system $\mathbb{S}$ in the state $|\Psi\rangle$, and measurement of the 
subsystem $A$ by some apparetus $M$;
\newline
(b) preparation of the system $\mathbb{S}'$ in the state $|\Psi'\rangle$, followed by measurement of the 
part $A$;
\newline
(c) the same as the experiment $a$, but, simultaneously with preparation of $\mathbb{S}$, the system $C'$ is 
prepared in some pure state $|\psi_{C'}\rangle$;
\newline
(d) the same as the experiment $b$, with preparation of the system $C$ in some pure state $|\psi_{C}\rangle$ 
simultaneously with preparation of $\mathbb{S}'$. 
\newline
In the experiment $d$, particles $C$ and $C'$ are swapped after preparation, 
in order to get the same configuration of particles as in the experiment $c$.

% \addtocounter{figure}{2}
\begin{figure}
\includegraphics[width=\linewidth]{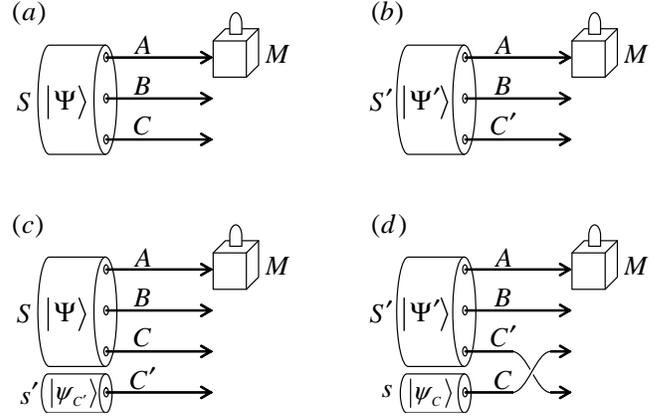}
\caption{Four thought experiments discussed in Appendix C. 
In experiments $a$ and $c$, the source $S$ emits particles $A$, $B$ and $C$ prepared in some joint pure state $|\Psi\rangle$. 
Similarly, in experiments $b$ and $d$ the source $S'$ emits particles $A$, $B$ and $C'$ prepared in some 
state $|\Psi'\rangle$. 
Additionally, in the last two experiments, the source $s$ ($s'$) emits a particle $C$ ($C'$) prepared 
in a pure state $|\psi_{C}\rangle$ ($|\psi_{C'}\rangle$). In experiments with two sources, both of them work simultaneously. 
Some time after the emission, the particle $A$ reaches the measuring apparatus $M$. The lamp on the apparatus $M$ indicates 
whether the measurement outcome is equal to some chosen value.
}
\label{fig:3}
\end{figure}

Let $\hat\rho_a$, $\hat\rho_b$, $\hat\rho_c$ and $\hat\rho_d$ be partial density matrices of the subsystem $A$ 
in experiments $a,b,c,d$. Obviously,
\begin{equation} \label{c:1}
  \hat\rho_a = \hat\rho_c ,  \quad  \hat\rho_b = \hat\rho_d .
\end{equation}
Then, let $P_a$, $P_b$, $P_c$ and $P_d$ be probabilities of some definite (chosen once and for all) outcome of measurement 
in experiments $a,b,c,d$, correspondingly. One can readily conclude that 
\begin{equation} \label{c:2}
  P_a = P_c .
\end{equation}
Indeed, the system $C'$ does not interact with the system $\mathbb{S}$, so any action with $C'$ 
(creation, preparation in some state, etc.) cannot alter probabilities of events, in which the system 
$\mathbb{S}$ (but not $C'$) is involved. The same argument shows that 
\begin{equation} \label{c:3}
  P_b = P_d .
\end{equation}

Now let us compare experiments $c$ and $d$. In both of them, the composite system $A+B+C+C'$ is in a pure state 
before the measurement of the part $A$. One can therefore repeat all the reasoning of the main part of this paper, 
implying that the part $B+C+C'$ serves as an environment (instead of the particle $B$ of the main part of this paper). 
As a result, one can conclude that if the partial density matrices of the part $A$ before the measurement 
are the same in both experiments $(\hat\rho_c = \hat\rho_d)$, then the probabilities of the chosen outcome are also 
the same $(P_c = P_d)$:
\begin{equation} \label{c:4}
  \hat\rho_c = \hat\rho_d  \; \Rightarrow \;  P_c = P_d .
\end{equation}

Finally, combining Eqs.~(\ref{c:1})--(\ref{c:4}), one can get 
\begin{equation} \label{c:5}
  \hat\rho_a = \hat\rho_b  \; \Rightarrow \;  P_a = P_b .
\end{equation}
Eq.~(\ref{c:5}) generalizes the statement of the main part of the paper that the probability of any measurement outcome 
depends on a (pure) state of a combined system ``measured object + environment'' only through the partial density matrix 
of the measured object. Now it is proven that the dependence of the probability $P$ on the density matrix $\hat\rho$ 
is universal with respect to choice of an environment.

\subsection{Appendix D. Proof of the equality $a_\lambda=\lambda$}

Let us consider two real-valued functions: a function $F(\hat\rho)$ 
of the density matrix $\hat\rho$ of some quantum system, 
and a function $a_\lambda$ of a real argument $\lambda\in[0,1]$. 
These functions are supposed to obey the following relation:
\begin{equation} \label{d:f-linearity-a-lambda}
  F \bigl( (1-\lambda)\hat\rho_0 + \lambda\hat\rho_1 \bigl) = 
  (1-a_\lambda) \, F (\hat\rho_0) + a_\lambda \, F (\hat\rho_1) ,
\end{equation}
that is valid for any density matrices $\hat\rho_0$, $\hat\rho_1$ 
and for any values of $\lambda\in[0,1]$. 
The function $F(\hat\rho)$ is not a constant.
The function $a_\lambda$ satisfies the following conditions:
\begin{gather} 
  \label{d:a-lambda-restriction}
  0 \leq a_\lambda \leq 1 , \\
  \label{d:a-lambda-01}
  a_0=0 , \quad a_1=1 .
\end{gather}
In this Section, we will prove that
\begin{equation} \label{d:a-lambda-is-lambda}
  \forall \lambda\in[0,1]  \quad  a_\lambda = \lambda .
\end{equation}

{\bf 1.} Let us substitute to Eq.~(\ref{d:f-linearity-a-lambda}) $\hat\rho_1$ as $\hat\rho_0$, $\hat\rho_0$ as $\hat\rho_1$, 
and $1-\lambda$ as $\lambda$. The result is
\begin{equation} \label{d1:inverse}
  F \bigl( \lambda\hat\rho_1 + (1-\lambda)\hat\rho_0 \bigl) = 
  (1-a_{1-\lambda}) \, F (\hat\rho_1) + a_{1-\lambda} \, F (\hat\rho_0) .
\end{equation}
Left-hand sides of Eqs.~(\ref{d:f-linearity-a-lambda}) and~(\ref{d1:inverse}) are the same. Subtracting right-hand sides 
one from another, one can get
\[
  (1-a_\lambda-a_{1-\lambda}) \left[ F(\hat\rho_0)-F(\hat\rho_1) \right] = 0 .
\]
Since one can choose such $\hat\rho_0$ and $\hat\rho_1$ that $F(\hat\rho_0) \neq F(\hat\rho_1)$, then 
$1-a_\lambda-a_{1-\lambda}=0$, i.~e. 
\[
  \forall \lambda\in[0,1]  \quad  a_\lambda + a_{1-\lambda} = 1 .
\]
In particular, $a_{1/2} + a_{1/2} = 1$, i.~e.
\[
  a_{1/2} = 1/2 .
\]

{\bf 2.} Let us introduce a shorthand notation $\hat\rho_\lambda$,
\[
  \hat\rho_\lambda  \stackrel{\text{def}}{=}  (1-\lambda)\hat\rho_0 + \lambda\hat\rho_1 ,
\]
and write Eq.~(\ref{d:f-linearity-a-lambda}) for $\lambda=x$, $\lambda=y$, and $\lambda=(x+y)/2$, where $x$ and $y$ 
are some real numbers between 0 and 1:
\begin{gather}
  \label{d2:x}
  F(\hat\rho_x) = (1-a_x) \, F(\hat\rho_0) + a_x \, F(\hat\rho_1) , \\
  \label{d2:y}
  F(\hat\rho_y) = (1-a_y) \, F(\hat\rho_0) + a_y \, F(\hat\rho_1) , \\
  \label{d2:xy}
  F(\hat\rho_{(x+y)/2}) = (1-a_{(x+y)/2}) \, F(\hat\rho_0) + a_{(x+y)/2} \, F(\hat\rho_1) .
\end{gather}
On the other hand, the matrix $\hat\rho_{(x+y)/2}$ is a linear combination of matrices $\hat\rho_x$ and $\hat\rho_y$:
\[
  \hat\rho_{(x+y)/2} = \frac12 \, \hat\rho_x + \frac12 \, \hat\rho_y .
\]
Hence, Eq.~(\ref{d:f-linearity-a-lambda}) for $\hat\rho_x$ as $\hat\rho_0$, $\hat\rho_y$ as $\hat\rho_1$, and 
$1/2$ as $\lambda$ gives
\begin{equation} \label{d2:xy2}
  F(\hat\rho_{(x+y)/2}) = (1-a_{1/2}) \, F(\hat\rho_x) + a_{1/2} \, F(\hat\rho_y) .
\end{equation}
Substituting expressions for $F(\hat\rho_x)$, $F(\hat\rho_y)$ and $F(\hat\rho_{(x+y)/2})$ 
from Eqs.~(\ref{d2:x})--(\ref{d2:xy}) into Eq.~(\ref{d2:xy2}), and taking into 
account that $a_{1/2} = 1/2$, one can get
\begin{gather*}
  (1-a_{(x+y)/2}) \, F(\hat\rho_0) + a_{(x+y)/2} \, F(\hat\rho_1) = \\
  = \frac12 \left[ (1-a_x) \, F(\hat\rho_0) + a_x \, F(\hat\rho_1) \right] \\
  + \frac12 \left[ (1-a_y) \, F(\hat\rho_0) + a_y \, F(\hat\rho_1) \right] .
\end{gather*}
Let us subtract $F(\hat\rho_0)$ from both sides of the last equation, and get the following:
\[
  a_{(x+y)/2} \left[ F(\hat\rho_1) - F(\hat\rho_0) \right] = 
  \frac{a_x+a_y}{2} \, \left[ F(\hat\rho_1) - F(\hat\rho_0) \right] .
\]
Since the matrices $\hat\rho_0$ and $\hat\rho_1$ can be chosen such that $F(\hat\rho_1) - F(\hat\rho_0) \neq 0$, then
\[
  a_{(x+y)/2} = \frac{a_x+a_y}{2} .
\]

{\bf 3.} Let us consider the matrix $\hat\rho_{xy}$. On the one hand, one can write a relation analogous to 
Eq.~(\ref{d2:x}),
\begin{equation} \label{d3:xy}
  F(\hat\rho_{xy}) = (1-a_{xy}) \, F(\hat\rho_0) + a_{xy} \, F(\hat\rho_1) .
\end{equation}
On the other hand, the matrix $\hat\rho_{xy}$ can be expressed via $\hat\rho_0$ and $\hat\rho_x$:
\[
  \hat\rho_{xy} = (1-y) \, \hat\rho_0 + y \, \hat\rho_x ,
\]
therefore,
\begin{equation} \label{d3:xy2}
  F(\hat\rho_{xy}) = (1-a_y) \, F(\hat\rho_0) + a_y \, F(\hat\rho_x) .
\end{equation}
Then, we substitute $F(\hat\rho_x)$ and $F(\hat\rho_{xy})$ from Eqs.~(\ref{d2:x}), (\ref{d3:xy}) into Eq.~(\ref{d3:xy2}):
\begin{multline*}
  (1-a_{xy}) \, F(\hat\rho_0) + a_{xy} \, F(\hat\rho_1) = \\
  = (1-a_y) \, F(\hat\rho_0) + a_y \, \left[ (1-a_x) \, F(\hat\rho_0) + a_x \, F(\hat\rho_1) \right] ,
\end{multline*}
subtract $F(\hat\rho_0)$ from both sides:
\[
  a_{xy} \left[ F(\hat\rho_1) - F(\hat\rho_0) \right] = 
  a_x a_y \left[ F(\hat\rho_1) - F(\hat\rho_0) \right] ,
\]
and divide both sides by $\left[ F(\hat\rho_1) - F(\hat\rho_0) \right]$. The result is:
\[
  a_{xy} = a_x a_y .
\]

{\bf 4.} Let us summarize what is known about the function $a_\lambda$ up to now:
\begin{equation} \label{d4:1}
  a_0=0; \quad a_1=1; \quad a_{1/2}=1/2;
\end{equation}
for any $x\in[0,1]$
\begin{gather}
  \label{d4:2}
  0 \leq a_x \leq 1 ; \\
  \label{d4:3}
  a_x + a_{1-x} = 1 ;
\end{gather}
for any $x\in[0,1]$ and $y\in[0,1]$
\begin{gather}
  \label{d4:4}
  a_{(x+y)/2} = \frac{a_x+a_y}{2} \, ; \\
  \label{d4:5}
  a_{xy} = a_x a_y .
\end{gather}

{\bf 5.} Using Eq.~(\ref{d4:4}) repeatedly, one can calculate the function $a_\lambda$ for infinitely many 
values of $\lambda$, as follows:
\begin{align*}
  a_{1/4} &= a_{(0+1/2)/2} = \frac{ a_{0}+a_{1/2} }2 = \frac{ 0+1/2 }2 = 1/4 , \\
  a_{3/4} &= a_{(1/2+1)/2} = \frac{ a_{1/2}+a_{1} }2 = \frac{ 1/2+1 }2 = 3/4 , \\
  a_{1/8} &= a_{(0+1/4)/2} = \frac{ a_{0}+a_{1/4} }2 = \frac{ 0+1/4 }2 = 1/8 , \\
  a_{3/8} &= a_{(1/4+1/2)/2} = \frac{ a_{1/4}+a_{1/2} }2 = \frac{ 1/4+1/2 }2 = 3/8 , \\
  a_{5/8} &= a_{(1/2+3/4)/2} = \frac{ a_{1/2}+a_{3/4} }2 = \frac{ 1/2+3/4 }2 = 5/8 , 
\end{align*}
and so on. The result is $a_\lambda = \lambda$ for rational values $\lambda=p/2^q$, where $q=1,2,3,\ldots,$ and 
$p=0,1,2,\ldots,2^q$.

{\bf 6.} Now let us prove that for any $x\in[0,1]$ and $y\in[0,1]$
\begin{equation} \label{d6}
  x \leq y   \; \Rightarrow \;   a_x \leq a_y .
\end{equation}
For this, we write $x$ as $t y$, where $t\in[0,1]$. From Eq.~(\ref{d4:5}), 
\[
  a_x = a_t a_y .
\]
From Eq.~(\ref{d4:2}),
\[
  a_t \leq 1 .
\]
Hence, $a_x \leq a_y$, and Eq.~(\ref{d6}) is proven.

{\bf 7.} Consider now an arbitrary real number $\lambda\in[0,1]$. Let us construct a series 
$\lambda_1,\lambda_2,\ldots$ of rational numbers by the following rule:
\[
  \lambda_q = \frac{[2^q\lambda]}{2^q} ,
\]
where square brackets denote taking the integer part.
By construction, 
\begin{equation} \label{d7:lambda-vs-lambda-q}
  \forall q  \quad  \lambda \geq \lambda_q ,
\end{equation}
and the difference $\lambda - \lambda_q$ goes to zero when $q$ is growing. So,
\begin{equation} \label{d7:lambda-sup}
  \lambda = \sup \, \{\lambda_q\} .
\end{equation}
Since the numbers $\lambda_q$ have the form of $p/2^q$, then
\[
  a_{\lambda_q} = \lambda_q .
\]
On the other hand, due to Eqs.~(\ref{d6}) and (\ref{d7:lambda-vs-lambda-q}) 
\[
  a_\lambda \geq a_{\lambda_q} .
\]
Therefore
\[
  \forall q  \quad  a_\lambda \geq \lambda_q ,
\]
which means
\begin{equation} \label{d7:a-lambda-sup}
  a_\lambda \geq \sup \, \{\lambda_q\} .
\end{equation}
Comparing Eqs.~(\ref{d7:lambda-sup}) and~(\ref{d7:a-lambda-sup}), one can see that 
\begin{equation} \label{d7:a-lambda-vs-lambda}
  \forall \lambda\in[0,1]  \quad  a_\lambda \geq \lambda .
\end{equation}

{\bf 8.} Finally, Eq.~(\ref{d:a-lambda-is-lambda}) can be proven by contradiction. 
Let Eq.~(\ref{d:a-lambda-is-lambda}) is wrong, i.~e. there is such a number $\lambda\in[0,1]$ that 
$a_\lambda \neq \lambda$. Then, according to Eq.~(\ref{d7:a-lambda-vs-lambda}),
\begin{equation} \label{d8:1}
  a_\lambda > \lambda .
\end{equation}
Also, according to Eq.~(\ref{d7:a-lambda-vs-lambda}),
\begin{equation} \label{d8:2}
  a_{1-\lambda} \geq 1-\lambda .
\end{equation}
Adding Eq.~(\ref{d8:1}) and Eq.~(\ref{d8:2}), one can get
\begin{equation} \label{d8:3}
  a_\lambda + a_{1-\lambda} > 1 ,
\end{equation}
which contradicts to Eq.~(\ref{d4:3}). 
Thus, Eq.~(\ref{d:a-lambda-is-lambda}) is proven.

\subsection{Appendix E. Trace form for any linear function of density matrix}

Let $\hat\rho$ denote a density matrix of some quantum system having the $N$-dimensional state space. 
In other words, $\hat\rho$ denotes a non-negative Hermitial matrix $N\times N$ with unit trace. 
Such density matrix can be parametrized by $N^2-1$ real numbers: 
$N-1$ diagonal matrix elements $\rho_{11},\ldots,\rho_{N-1,N-1}$; $N(N-1)/2$ real parts of non-diagonal 
elements $\rho_{mn}$, $m<n$; and $N(N-1)/2$ imaginary parts of these non-diagonal elements.
Then, any function of density matrix can be considered as a function of $N^2-1$ real arguments 
listed above.

Let $F(\hat\rho)$ be a real-valued function of density matrix, and it is linear on $N^2-1$ real parameters 
$\rho_{11},\ldots,\rho_{N-1,N-1}$, $\text{Re}\,\rho_{mn}$, $\text{Im}\,\rho_{mn}$. 
In this Section, we will demonstrate that any such linear function can be represented as
\begin{equation} \label{e:f-trace}
  F(\hat\rho) = \mathrm{Tr} (\hat A \hat\rho) , 
\end{equation}
where $\hat A$ is an Hermitian matrix $N\times N$, and will find its matrix elements $A_{mn}$.

First, we will write down the function $F(\hat\rho)$ explicitly, using its linearity:
\begin{equation} \label{e:f-explicit}
  F(\hat\rho) = a + \sum_{n=1}^{N-1} b_n \rho_{nn} 
  + \sum_{m<n} c_{mn} \text{Re}\,\rho_{mn}
  + \sum_{m<n} d_{mn} \text{Im}\,\rho_{mn} ,
\end{equation}
where $a,b_n,c_{mn},d_{mn}$ are some coefficients. One can find $N$ coefficients $a,b_1,\ldots,b_{N-1}$ 
from values of the function $F$ for density matrices corresponding to the basis vectors 
$|1\rangle,\ldots,|N\rangle$:
\begin{align}
  \label{e:f-abcd1}
  & F(|n\rangle \langle n|) = a + b_n \text{ for } n=1,\ldots,N-1 ; \\
  \label{e:f-abcd2}
  & F(|N\rangle \langle N|) = a .
\end{align}
The coefficients $c_{mn},d_{mn}$ $(m<n)$ can be expressed as follows:
\begin{gather}
  \label{e:f-abcd3}
  c_{mn} = \frac{\partial F(\hat\rho)}{\partial\, \text{Re}\,\rho_{mn}} \, , \\
  \label{e:f-abcd4}
  d_{mn} = \frac{\partial F(\hat\rho)}{\partial\, \text{Im}\,\rho_{mn}} \, .
\end{gather}

Then, consider an expression
\begin{equation} \label{e:tr}
  \mathrm{Tr} (\hat A \hat\rho)  \equiv  \sum_{m=1}^N \sum_{n=1}^N A_{mn} \rho_{nm} ,
\end{equation}
in which $\hat A$ is an Hermitian matrix $N\times N$. Let us rewrite this expression in a form 
similar to Eq.~(\ref{e:f-explicit}). For this, we separate diagonal terms from non-diagonal ones:
\begin{equation} \label{e:tr-separate}
  \mathrm{Tr} (\hat A \hat\rho) 
  = \sum_{n=1}^N A_{nn}\rho_{nn} 
  + \sum_{m<n} (A_{mn}\rho_{nm} + A_{nm}\rho_{mn}) .
\end{equation}
Then, we get rid of the matrix element $\rho_{NN}$, expressing it via the rest diagonal elements,
\begin{equation} \label{e:rho-nn}
  \rho_{NN} = 1 - \sum_{n=1}^{N-1} \rho_{nn} .
\end{equation}
Using Eq.~(\ref{e:rho-nn}), one can write the first sum of Eq.~(\ref{e:tr-separate}) in the form
\begin{equation} \label{e:tr-first}
  \sum_{n=1}^N A_{nn}\rho_{nn} = A_{NN} + \sum_{n=1}^{N-1} (A_{nn}-A_{NN}) \rho_{nn} .
\end{equation}
Each term of the second sum in Eq.~(\ref{e:tr-separate}) can be rewritten as follows (taking into account that 
$A_{nm}=A_{mn}^*$ and $\rho_{nm}=\rho_{mn}^*$):
\begin{multline} \label{e:tr-nondiag}
  A_{mn}\rho_{nm} + A_{nm}\rho_{mn} = \\
  = 2\, \text{Re}\,A_{mn} \text{Re}\,\rho_{mn} 
  + 2\, \text{Im}\,A_{mn} \text{Im}\,\rho_{mn} .
\end{multline}
Substitution of Eqs.~(\ref{e:tr-first}),~(\ref{e:tr-nondiag}) into Eq.~(\ref{e:tr-separate}) gives
\begin{multline} \label{e:tr-similar}
  \mathrm{Tr} (\hat A \hat\rho) = 
  A_{NN} + \sum_{n=1}^{N-1} (A_{nn}-A_{NN})\, \rho_{nn} \\
  + \sum_{m<n} 2\, \text{Re}\,A_{mn} \text{Re}\,\rho_{mn} 
  + \sum_{m<n} 2\, \text{Im}\,A_{mn} \text{Im}\,\rho_{mn} .
\end{multline}

Comparing Eq.~(\ref{e:f-explicit}) with Eq.~(\ref{e:tr-similar}), one can conclude that the functions 
$F(\hat\rho)$ and $\mathrm{Tr} (\hat A \hat\rho)$ will coinside for all $\hat\rho$, if the coefficients 
$a,b,c,d$ are
\begin{align*}
  & a      = A_{NN} , \\
  & b_n    = A_{nn} - A_{NN} , \\
  & c_{mn} = 2\,\text{Re}\,A_{mn} , \\
  & d_{mn} = 2\,\text{Im}\,A_{mn} .
\end{align*}
Using these relations, one can fully define the matrix $\hat A$ in terms of the coefficients $a,b,c,d$:
\begin{align}
  \label{e:a-abcd1}
  & A_{NN} = a , \\
  \label{e:a-abcd2}
  & A_{nn} = b_n + a \quad (\text{for } n=1,\ldots,N-1) , \\
  \label{e:a-abcd3}
  & A_{mn} = \frac{c_{mn}+id_{mn}}2 \quad (\text{for } m<n) .
\end{align}
Finally, let us derive the values of coefficients $a,b,c,d$ from Eqs.~(\ref{e:f-abcd1})--(\ref{e:f-abcd4}) 
and substitute these values into Eqs.~(\ref{e:a-abcd1})--(\ref{e:a-abcd3}). As a result, 
diagonal matrix elements $A_{nn}$ ($n=1,\ldots,N$) are
\begin{equation} \label{e:a-diagonal}
  A_{nn} = F(|n\rangle \langle n|) ,
\end{equation}
and non-diagonal elements $A_{mn},A_{nm}$ ($m<n$) are
\begin{equation} \label{e:a-non-diagonal}
  A_{mn} = (A_{mn})^* 
  = \frac 12\, \frac{\partial F(\hat\rho)}{\partial\, \text{Re}\,\rho_{mn}} 
  + \frac i2\, \frac{\partial F(\hat\rho)}{\partial\, \text{Im}\,\rho_{mn}} \, .
\end{equation}

Thus, it is shown that Eq.~(\ref{e:f-trace}) is valid for all density matrices $\hat\rho$, 
if the Hermitian matrix $\hat A$ is chosen according to Eqs.~(\ref{e:a-diagonal}) and~(\ref{e:a-non-diagonal}).

\subsection{Appendix F. Born rule from POVM}

In this Section, we take for granted that any measurement in quantum mechanics is described by a POVM, i.~e. 
for each ($k$th) outcome of a measurement performed by an apparatus $M$, 
there is such an Hermitian operator $\hat A_M^{(k)}$ that the probability $P_M^{(k)}$ of this outcome is 
\begin{equation} \label{f:povm}
  P_M^{(k)} = \mathrm{Tr} (\hat A_M^{(k)} \hat\rho) ,
\end{equation}
where $\hat\rho$ is the density matrix of the measured system before the measurement. It follows from inequalities 
$0 \leq P_M^{(k)} \leq 1$ that all eigenvalues of the operator $\hat A_M^{(k)}$ are bound within the range 
$[0,1]$.

Let us derive the Born rule from Eq.~(\ref{f:povm}). We will consider the case of $maximal$ measurement, 
for which the number of possible outcomes is equal to the dimensionality $N$ of 
the measured system's state space.
Suppose that there is a set $\{S_1,S_2,\ldots,S_N\}$ of $N$ states, each of them $(S_k)$
yielding the \emph{definite} ($k$th) result of measurement by the apparatus $M$ with probability 1:
\begin{equation} \label{f:ideal-measurement}
  \forall k \quad P_M^{(k)} (S_k) = 1 . 
\end{equation}
One can conclude from Eq.~(\ref{f:povm}) and from the 
equality $\mathrm{Tr}(\hat\rho)=1$, that the value of $P_M^{(k)}$ cannot be larger than the largest 
eigenvalue of the operator $\hat A_M^{(k)}$. On the other hand, eigenvalues of $\hat A_M^{(k)}$ are bounded 
within the range $[0,1]$. Hence, Eq.~(\ref{f:ideal-measurement}) implies that at least one eigenvalue 
of $\hat A_M^{(k)}$ is equal to 1. Let a unit vector $|\varphi_k\rangle$ be the corresponding eigenvector. Then,
\[
  P_M^{(k)} (|\varphi_k\rangle) = \mathrm{Tr} (\hat A_M^{(k)} |\varphi_k\rangle \langle\varphi_k|) 
  = \langle\varphi_k| \hat A_M^{(k)} |\varphi_k\rangle = 1 .
\]
Since $\sum_{l=1}^N P_M^{(l)} (|\varphi_k\rangle) = 1$, then 
\[
  P_M^{(k)} (|\varphi_l\rangle) = 0 \text{ if } k \neq l .
\]
It follows from the latter equation and from non-negativity of the operator $\hat A_M^{(k)}$, that 
$|\varphi_l\rangle$ is the eigenvector of $\hat A_M^{(k)}$ with zero eigenvalue. If eigenvectors of 
an Hermitian operator correspond to different eigenvalues, they are mutually orthogonal. So 
\[
  \langle\varphi_k| \varphi_l\rangle = 0 \text{ if } k \neq l .
\]
Thus, the vectors $|\varphi_1\rangle,\ldots,|\varphi_N\rangle$ form an orthonormal basis in the $N$-dimensional 
state space. Each of $N$ operators $\hat A_M^{(1)},\ldots,\hat A_M^{(N)}$ is diagonalized in this basis:
\[
  \hat A_M^{(k)} |\varphi_l\rangle = \delta_{kl} |\varphi_l\rangle .
\]
Therefore, each operator $\hat A_M^{(k)}$ is actually a projector:
\[
  \hat A_M^{(k)} = |\varphi_k\rangle \langle\varphi_k| ,
\]
whence
\[
  P_M^{(k)} = \langle\varphi_k| \hat\rho |\varphi_k\rangle .
\]
It can be seen from this equation that the probability $P_M^{(k)}$ reaches 1 only for the pure state with 
the wavefunction $|\varphi_k\rangle$. Therefore all the states $S_k$ are pure, and their state vectors are 
mutually orthogonal.

In the case of an arbitrary pure state $|\psi\rangle$ of the measured system,
\[
  P_M^{(k)} (|\psi\rangle) = \langle\varphi_k| (|\psi\rangle \langle\psi|) |\varphi_k\rangle 
  \equiv  \bigl| \langle\varphi_k|\psi\rangle \bigr|^2 .
\]
This is the Born rule.

% --------------------------

\bibliography{envariance}

\end{document}